\documentclass[sigconf]{acmart}

\AtBeginDocument{%
  \providecommand\BibTeX{{%
    \normalfont B\kern-0.5em{\scshape i\kern-0.25em b}\kern-0.8em\TeX}}}
\usepackage{listings, xcolor}
\lstdefinelanguage{gta}{
    numbers=left,
    numbersep=1em,
    numberstyle=\footnotesize\color{gray},
    alsoletter={:},
    keywords={name:,test:,next:,precmd:,postcmd:},keywordstyle=\color{red},
    moredelim=[s][\color{purple}]{\{\{}{\}\}}
}
\definecolor{lightgray}{gray}{0.9}
\setcopyright{acmcopyright}
\copyrightyear{2021}
\acmYear{2021}
\setcopyright{acmlicensed}\acmConference[ITiCSE 2021]{26th ACM Conference on Innovation and Technology in Computer Science Education V. 1}{June 26-July 1, 2021}{Virtual Event, Germany}
\acmBooktitle{26th ACM Conference on Innovation and Technology in Computer Science Education V. 1 (ITiCSE 2021), June 26-July 1, 2021, Virtual Event, Germany}
\acmPrice{15.00}
\acmDOI{10.1145/3430665.3456387}
\acmISBN{978-1-4503-8214-4/21/06}

\usepackage{courier}

\newcommand\YAMLcolonstyle{\color{red}\mdseries}
\newcommand\YAMLkeystyle{\color{black}\bfseries\small\ttfamily}
\newcommand\YAMLvaluestyle{\color{blue}\mdseries}

\makeatletter

\newcommand\language@yaml{yaml}

\expandafter\expandafter\expandafter\lstdefinelanguage
\expandafter{\language@yaml}
{
  keywords={true,false,null,y,n},
  keywordstyle=\color{darkgray}\bfseries,
  basicstyle=\YAMLkeystyle,                                 
  sensitive=false,
  comment=[l]{\#},
  morecomment=[s]{/*}{*/},
  commentstyle=\color{purple}\ttfamily,
  stringstyle=\YAMLvaluestyle\ttfamily,
  moredelim=[l][\color{orange}]{\&},
  moredelim=[l][\color{magenta}]{*},
  moredelim=**[il][\YAMLcolonstyle{:}\YAMLvaluestyle]{:},   
  morestring=[b]',
  morestring=[b]",
  literate =    {---}{{\ProcessThreeDashes}}3
                {>}{{\textcolor{red}\textgreater}}1     
                {|}{{\textcolor{red}\textbar}}1 
                {\ -\ }{{\mdseries\ -\ }}3,
}

\lst@AddToHook{EveryLine}{\ifx\lst@language\language@yaml\YAMLkeystyle\fi}
\makeatother
\usepackage{bm}

\begin{document}

\title{TermAdventure: Interactively Teaching UNIX Command Line, Text Adventure Style}

\author{Marek \v{S}uppa}
\authornote{The authors contributed equally to this research.}
\email{marek@suppa.sk}
\orcid{0000-0002-5887-0696}
\affiliation{%
  \institution{Comenius University}
  \city{Bratislava}
  \country{Slovakia}
}

\author{Ondrej Jariabka}
\authornotemark[1]
\email{o.jariabka@gmail.com}
\affiliation{%
  \institution{Comenius University}
  \city{Bratislava}
  \country{Slovakia}
}

\author{Adri\'{a}n Matejov}
\authornotemark[1]
\email{ado.matejov@gmail.com}
\affiliation{%
  \institution{Comenius University}
  \city{Bratislava}
  \country{Slovakia}
}

\author{Marek Nagy}
\email{mnagy@ii.fmph.uniba.sk}
\affiliation{%
  \institution{Comenius University}
  \city{Bratislava}
  \country{Slovakia}
}

\renewcommand{\shortauthors}{\v{S}uppa and Jariabka, et al.}

\begin{abstract}
  Introductory UNIX courses are typically organized as lectures, accompanied by a set of exercises, whose solutions are submitted to and reviewed by the lecturers. While this arrangement has become standard practice, it often  requires the use of an external tool or interface for submission and does not automatically check its correctness. That in turn leads to increased workload and makes it difficult to deal with potential plagiarism.
  
  In this work we present TermAdventure (TA), a suite of tools for creating interactive UNIX exercises. These resemble text adventure games, which immerse the user in a text environment and let them interact with it using textual commands. In our case the ''adventure'' takes place inside a UNIX system and the user interaction happens via the standard UNIX command line. The adventure is a set of exercises, which are presented and automatically evaluated by the system, all from within the command line environment. The suite is released under an open source license, has minimal dependencies and can be used either on a UNIX-style server or a desktop computer running any major OS platform through Docker.
  
  We also reflect on our experience of using the presented suite as the primary teaching tool for an introductory UNIX course for Data Scientists and discuss the implications of its deployment in similar courses. The suite is released under the terms of an open-source license at \url{https://github.com/NaiveNeuron/TermAdventure}.
  
\end{abstract}




\maketitle

\section{Introduction}

  Given its ubiquitous presence in modern computing, the ability to effectively make use of UNIX systems is crucial for any of its users. These increasingly include not only the students of Computer Science and related disciplines (such as Mathematics) \cite{doyle2007teach} but also students of Natural and Social sciences.
  
  Although the fundamentals of UNIX have largely stayed the same over the past years, the students, who first experience it at the beginning of their higher studies, have changed significantly. Despite growing up in a world with a lot of technology and being sometimes referred to as ''digital natives'' \cite{gunther2007digital}, they often struggle to adapt to the command line environment of an operating system which they are likely not familiar with. As Kendon and Stephenson note in \cite{kendon2016unix}, this may be because \emph{''it is so removed from their prior computing exprience''}.
  
  The standard approach to introducing students to the command line entails a set of exercises, whose solutions are then submitted and later reviewed by the class staff. Such an approach introduces a time lag by not providing immediate feedback, which often leaves the students uncertain about the correctness of their solution, leads to frustration, increases workload of the class staff and makes it difficult to deal with potential plagiarism. Furthermore, with increasing size of classes it is becoming more difficult to get an overview of the submitted solutions or to notice when a student is stuck on a problem during the exercise.
  
  In this work we present TermAdventure (TA), a suite of tools that allows for interactive introduction to UNIX command line. It does so in the style of text adventure games, which immerse the user in a text environment and allow them to interact with it using various commands. In our case the ''adventure'' takes place inside a UNIX command line and the interaction happens using the standard UNIX commands.
  TermAdventure leads the user through the adventure, consisting of exercises (levels). In each level it presents the challenges and automatically tests the provided solutions. The levels are designed through the TermAdventure templating capabilities and supports randomization so that each student can be given an unique challenge.

  
  We further provide a discussion on using the presented system as a primary learning tool for an introductory UNIX command line course for Data Scientists, which took place in a remote fashion due to COVID-19 pandemic.

\begin{figure}
\centering
\includegraphics[width=0.5\textwidth]{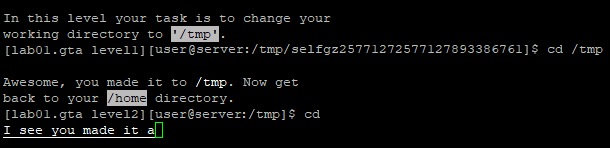}
\caption{A lab assignment through TermAdventure. The user is inside their shell session, which responds to their actions (i.e. moving to \texttt{/tmp} results in printing a congratulatory text).}
\label{pic:assigment}
\end{figure}

\section{TermAdventure}

Our inspiration to design and implement TermAdventure has come from personal experience of teaching an introductory UNIX command line class, in which the exercises were released in textual form via an e-learning system. The students would try to complete a specific exercise inside their command lines and then copy the commands leading to completion of the exercise back to the e-learning system. While this was straightforward to setup, there were numerous downsides to this arrangement. Namely, it was cumbersome for the students, who would often need to worry not only about actually solving the task in the exercise but also about posting the correct sequence of commands to the e-learning system. Furthermore, they would not know whether their solution was actually correct until the teaching staff manually review them.

Hence when designing TermAdventure, we had the following goals in mind:

\begin{description}
    \item[Real UNIX command line] Since the goal of the course is to make the students comfortable with the UNIX command line, it is imperative for the designed system not to require them to use any other tools. Additionally, this also means that keypress combinations, such as \texttt{Ctrl+w}, should work as well\footnote{In Bash and similar UNIX shells, \texttt{Ctrl+w} removes the last word currently written on the command line. In web browsers, however, it usually causes the current tab to close, and in turn renders a faithful Bash experience difficult to re-implement.}.
    
    \item[Immediate evaluation] The shorter the feedback loop, the higher the chance of the student walking away from an exercise with a feeling of accomplishment. In the best case scenario, the student should know right away when they managed to fulfil a task they worked on.
    
    
    \item[Plaintext configuration] The text of the respective assignments, as well as the relationships between them, should be stored in a plaintext file format. This ensures future compatibility and at the same time makes it easy for the assignments to be version tracked in version control systems.
    
    \item[Platform independence] It should be possible to run the assignments on a UNIX-style server as well as locally, whether on a UNIX system or a virtualized environment, such as a Virtual Machine or a Docker container.
    
    \item[Visibility for instructors] The instructors ought to be able to detect when a student struggles with a particular task. For example, this could be done by checking for limited interaction with the system in a specific timeframe or mindless trying of multiple commands.
    
\end{description}

Taking the goals above into account, we chose to implement TermAdventure as a Bash script accompanied by a single Go binary, whose sole purpose is to spawn a new, slightly altered Bash shell session, which communicates with the user (in our case the student) in a conversational style: it first prints out the description of the task and the user then uses standard Bash commands to fulfil it. The user experience this creates is detailed in Section \ref{sec:functionality-walkthrough}. The goal of providing visibility to instructors has been left to a separate part of the suite, called TermAdventure Monitor, which we describe in greater detail in Section \ref{sec:gta-monitor}.

\subsection{Functionality Walk-through}
\label{sec:functionality-walkthrough}

One of the major advantages of TermAdventure is that it runs directly inside user's command line environment and therefore removes the disconnect between user and the command line as they are not required to switch between different tools. The user is completing entire ''adventure'' inside the UNIX environment and becomes familiar with its behaviour while solving the presented problems.

Once the user starts the adventure, TermAdventure presents the problem statement by printing it character-by-character with a small delay (on the order of 50 milliseconds). The end-of-sentence characters (\texttt{.}, \texttt{!} and \texttt{?}) impose a larger delay (on the order of 500 milliseconds), which creates a dramatic effect, intrinsic to ''text adventures''.  This behaviour can be skipped by user by pressing Enter or Space keys, or removed altogether. The printing routine supports Bash formatting capabilities (bold, italic, underscore, similar to Markdown), which are visualized in Figure \ref{pic:assigment}.

Once the entire problem statement is presented, user control is returned to their normal command line environment, where they can use any available tools to solve the problem. Each executed user command is evaluated by a predefined test. The test is executed as a \texttt{Bash} command, located on the OS where TermAdventure is running. The fact the user passed a given challenge is  determined by test's return value. If the test passes, the user continues straight to the next challenge where he is again presented with text of the following problem. If the test does not pass, the user stays on the same challenge until they successfully solve the problem. This \textit{instant feedback} on the correctness of the solution is the second main advantage of TermAdventure. Without the feedback students often found a partial solution that seemed right but could not be assigned the full credit, which inevitably resulted in student frustration. Furthermore, since the explanation of why the solution is wrong came later, students found it difficult to remember the mistakes in their solutions. When tested again for the same problem, often times the same mistake was made.

While working on a specific assignment, the challenge description can easily get scrolled out of one's terminal, which would be full of output of other commands executed in the meantime. To bring back the description of the current task, users can execute the \texttt{ta\_print\_again} command which prints the task description to standard output. Moreover, the up-to-date task description can also be found in their home directory, specifically in \texttt{\$HOME/ta\_current\_level.txt}. Finally, if the user gets stuck, they can simply execute the \texttt{ta\_help} command. This will alert the instructors via the monitoring interface, which is described in greater detail in Section \ref{sec:gta-monitor}.

\subsection{Creating Challenges}
\label{sec:creating-challenges}

The full adventure usually consists of many small challenges (levels). Each level definition is specified in separate YAML \cite{ben2009yaml} file, that is read at the beginning of the adventure and levels are generated accordingly. The definition of a level consists of 
\begin{enumerate}
    \item \texttt{name} - Name of the level.
    \item \texttt{test} - Test for correctness of the solution.
    \item \texttt{next} - List of names of the possible next levels (level not pointing to any other level is considered to be the last).
    \item Textual description of the level's task.
\end{enumerate}

Using this setup, the levels of one adventure create a Directed Acyclic Graph (DAG), on which users flow until they reach any leaf. This branching strategy provides an easy way of creating different versions of the same level or different random paths for each student through the graph. It can lead to unique experience for each user and helps with potential plagiarism in a classroom setting.


To create an adventure, a user (in our case a lecturer) has to create a so called ''challenge definition'' file. It consists of all the textual description of the respective levels, which together comprise the DAG. Each level starts with the specification of metadata key-value pairs, followed by the text of the level separated by empty lines. The text is formatted using standard Markdown syntax. Levels are separated by the markdown horizontal line with dashes \texttt{(-----)}. Listing \ref{lst:sample_challenge} shows a sample ''challenge definition'' file, which creates a very simple, three-level adventure.

\begin{lstlisting}[language=gta, label={lst:sample_challenge}, caption=A sample ''challenge definition'' file. Note that text highlighting is done using standard markdown syntax.]
name: level1
next: ['level2']
test: /usr/bin/test $(pwd) = "/tmp"

In this level your task is to change your
working directory to *'/tmp'*.

--------------------

name: level2
next: ['finish']
test: /usr/bin/test $(pwd) = "$HOME"

Awesome, you made it to **/tmp**. Now get
back to your */home* directory.

--------------------

name: finish
test: false

# I see you made it again, awesome!
That's all for now, so lay back and enjoy
your shell!

--------------------
\end{lstlisting}

As previously mentioned, the \texttt{next} keyword specifies a \textit{list} of potential succeeding levels, one of which is randomly selected at runtime. This, together with different versions of the same level, creates multiple versions of the same adventure.

To alleviate the workload spent on creating the same level with just a few different words, TermAdventure also provides templating capabilities. Templates are created in the same way and format as challenge definition files. The template contains static parts, usually the majority of the text, as well as some special syntax describing how dynamic content will be inserted.
It utilizes Go's \texttt{html/templates} package as a template engine. This package is normally used for HTML templating, is therefore well tested and contains many features relevant for our use case (such as loops, variables or filters).
To create a template, build a normal challenge definition file, with the dynamic part written in standard Go template language, such as using \texttt{\{\{}, \texttt{\}\}} to mark dynamic variables.

Values for the template variables go to a separate yaml file containing key-value pairs, where the key is a name of the variable inside the template file. A simple template with its accompanying variable yaml file can be seen in Listing \ref{lst:sample_template} and \ref{lst:sample_template_vars}, respectively.
TermAdventure comes with few predefined filters, like \texttt{generate\_levels}, which is used to generate a list of upcoming levels for the \texttt{next} keyword as seen in Listing \ref{lst:sample_template}. It takes an iterable from the variables file as an argument together with the name format of the levels. For example, in Listing \ref{lst:sample_template} the output of the filter would be
\colorbox{lightgray}{\lstinline[basicstyle=\ttfamily]|['lvl21', 'lvl22', 'lvl23']|}.

All of these features create a simple and a straightforward way of creating a adventure, all of which can be done just from the command line itself. For this reason a TermAdventure challenge can run even on GUI-less environment, such as servers, and is very portable from one environment to the other. Furthermore, since the \textit{adventures} use only bash capabilities, such as the \texttt{test} command, once the adventure is created it can run on most of the modern UNIX architectures.

\begin{lstlisting}[language=gta, label={lst:sample_template}, caption=An example of a ''adventure'' template file.]
name: lvl1
next: [{{generate_levels "lvl2" $.dirs "%s%d"}}]
test: /usr/bin/test $(pwd) = "/tmp"

In this level your task is to change your
working directory to *'/tmp'*.

--------------------
{{ range $index, $dir := .directories }}
name: lvl2{{ add $index 1 }}
next: ['finish']
test: /usr/bin/test $(pwd) = "{{ dir }}"

Awesome, you made it to **/tmp**. Now get
to the */{{ dir }}* directory.

--------------------
{{ end }}
name: finish
test: false

# I see you made it again, awesome!
That's all for now, so lay back and enjoy
your shell!

--------------------
\end{lstlisting}
\begin{lstlisting}[language=yaml, label={lst:sample_template_vars}, caption=An example of a variables file. This will results in \texttt{lvl2} from Listing \ref{lst:sample_template} have three variants each redirecting users to a different folder.]
directories:
  - var
  - dev
  - home
\end{lstlisting}

\subsection{Implementation}
\label{sec:implementation}

Most UNIX shells are implemented as an interactive read-eval-print loop (REPL). The shell first reads the user's program, executes it and then prints a so called ''prompt'' for another command. An example of such prompt can be \colorbox{lightgray}{\lstinline[basicstyle=\ttfamily]|giles@nikola:~|}, where user \texttt{giles} is executing commands on a computer with hostname \texttt{nikola}.
Some modern UNIX shells (such as Bash) allow arbitrary programs to be executed as part of creating this prompt. This is normally done to enable its customization -- the prompt can for instance feature information such as the execution time of the previous command. In particular, Bash contains a special variable called \texttt{PROMPT\_COMMAND}, whose content gets executed just before the prompt is displayed. The POSIX standard \cite{josey2004open}, which many UNIX systems implement, further defines a specific \texttt{PS1} variable, which contains the prompt that is printed out when the shell is ready to read new commands. The standard approach to customizing the contents of the prompt therefore involves running arbitrary commands as part of \texttt{PROMPT\_COMMAND} and using their output to set the \texttt{PS1} variable to the desires value.

The core of the TermAdventure application is implemented as a single Go binary, which gets executed as one of the commands in \texttt{PROMPT\_COMMAND}. It reads a specific challenge file (such as the one outlined in Listing \ref{lst:sample_challenge}), and checks whether the prerequisites for the concrete level the user is in have been met.
The TermAdventure binary then parses the text of the level it just moved the user into, and prints it to the terminal, while implementing the Markdown-like formatting via ANSI escape sequences. For easier navigation, it also adds the name of the current challenge, as well as the level the user is currently on to \texttt{PS1}, which is also visualized in Figure \ref{pic:assigment}.

To set variables such as \texttt{PROMPT\_COMMAND} and \texttt{PS1}, the Bash shell is normally instructed to read a specific configuration file called \texttt{bashrc}. TermAdventure ships with a specific \texttt{bashrc} file, which injects it into \texttt{PROMPT\_COMMAND} and also ensures each user gets to work in the exact same environment.

To use TermAdventure, each user needs at least three files: the TermAdventure binary, the ''challenge definition'' and the specific \texttt{bashrc} file. To make running the challenges as user friendly as possible, we use an utility called \texttt{makeself}\footnote{\url{https://makeself.io/}}, which takes all the aforementioned files and wraps them into a single self-extractable compressed archive. The end user can then run it as any other shell script or application. This choice also supports future extensibility, as the compressed archive can be used to distribute various assets related to the assignment itself (i.e. text files to be formatted using standard UNIX text formatting tools).

Since the distribution via self-extractable archives only depends on the widely supported Bash, it should be possible to run it virtually on any UNIX-style server or desktop system (such as Linux, BSD or macOS), provided the TermAdventure binary gets recompiled for the target platform as well. On other systems (such as Windows), the resulting script can run inside a standard Docker container of any standard Linux distribution (e.g. Ubuntu).

\subsection{Security Implications}
\label{sec:security}

In order for TermAdventure to evaluate whether the user moved from one level to the other, it needs to know at which level is the user currently located. Since it only gets executed once in a while (before the shell prompt gets printed out), the user's current level needs to be saved to persistent storage. Saving this information to disk in plaintext would open it to a fairly trivial way of solving the assignment: the attacker would only need to find the place where TermAdventure saves its user's progress and change the current level to the final one.

To limit the viability of this attack, instead of saving the current level's name, TermAdventure stores its hash, which is computed as 

$$ hash = md5(salt_1 ~||~ \bm{challenge} ~||~ salt_2 ~||~ \bm{level} ~||~ salt_3 ~||~ \bm{home} ) $$

where $||$ denotes concatenation, $salt_{1,2,3}$ represent three separate parts of a salt, while $\bm{challenge}$, $\bm{level}$ and $\bm{home}$ denote the name of the TermAdventure challenge, the level in question and the user's home directory, respectively. In order to check what level the saved hash represents, TermAdventure iterates through the list of all levels it can find in the ''challenge file'', computes the aforementioned hash and compares it to the retrieved value.

The other attack vector to consider is the ''challenge definition'' file. It contains the command which decides whether the task in a particular level has been passed. While this does not directly unveil the solution in all cases, it provides a lot of information about the implementation of the exercise. To prevent it from being customarily revealed, it is encrypted using AES-128, AES-192, or AES-256 \cite{daemen1999aes}, depending on the length of the key, which is embedded in the binary.

Note that neither of these approaches are impenetrable, especially due to the open source nature of the whole suite. However, our aim here is to make it more difficult to reverse-engineer the assignment than to actually do it. Furthermore, thanks to the tools introduced in Section \ref{sec:gta-monitor}, any strange behavior, such as unsupported jumping between levels, should be easy to detect and react to.

\subsection{Testing}
\label{sec:testing}

While TermAdventure does some rudimentary sanity checks on the DAG that is built from the individual levels, mistakes inevitably happen. Consider for instance the test of the first level described in Listing \ref{lst:sample_challenge}. Swapping \texttt{/tmp} for \texttt{/temp} on a system that does not have a \texttt{/temp} directory would lead to a challenge that is impossible to finish.

\begin{lstlisting}[language=yaml, label={lst:yaml_test}, caption=A YAML file describing a sample test for the ''challenge description'' file introduced in Listing \ref{lst:sample_challenge}.]
executable: ./sample_challenge.sh
start_level: level1
finish_level: finish
tests:
  level1: cd /tmp
  level2: cd
\end{lstlisting}

In order to automatically prevent a situation like these from happening, TermAdventure ships with facilities for automatic testing of the various challenges. It is implemented by running a Bash session inside a Docker container, in which a TermAdventure challenge is running as well. A test consists of a set of commands associated with the respective levels in the challenge. A sample test can be found in Listing \ref{lst:yaml_test}. The evaluation software reads the test definition and starting from the first level (denoted as \texttt{start\_level}) executes commands according to the mapping in the \texttt{tests} dictionary. Each command is expected to move TermAdventure to the next level. When that does not happen, the test is deemed to fail, suggesting that a mistake may have happened either in the ''challenge definition'' file or the test itself. The procedure finishes when the test reaches the final level (denoted as \texttt{finish\_level}), meaning that at least one path of passing through the adventure DAG exists.

\section{TA Monitor}
\label{sec:gta-monitor}

TermAdventure itself does not provide either real time feedback or overview of all the solutions submitted. Such a feature is required if the class is visited by lots of students or when it is held remotely.

In order to provide such a real time feedback, we present
a tool called TA Monitor. It aims to continuously gather
solutions being submitted by students during the lab and provide an overview of what the class progress looks like at a given moment. It does so in real time and is capable of triggering notifications whenever a student is stuck at particular level. Moreover, once the lab is finished, it is able to automatically assign grades to the solutions and export the grades to third-party e-learning systems.

\begin{figure}[h]
\centering
\includegraphics[width=0.5\textwidth]{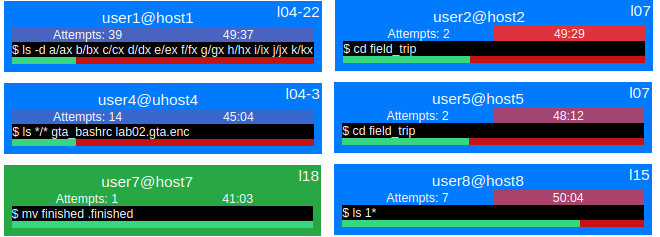}
\caption{Overview of the class provided by TA Monitor. Green boxes represent students that finished the exercise. Each box displays the current level (top right), last command used (bottom), the number of unsuccessful attempts and time since the last command execution.}
\label{pic:gta-monitor-class}
\end{figure}

The progress of the student is displayed in a small box, see Figure \ref{pic:gta-monitor-class}. The entire command history of the student's lab is shown after clicking on the box. The tool also provides lab statistics as well as a way to group similar solutions. The teacher is then
able to see which levels are the most problematic and
might be adjusted in the following editions of the course. All of this is displayed in a web browser, which once
again eliminates installing any other software.


\subsection{Implementation}
\label{sec:gta-monitor-implementation}

TA Monitor\footnote{While TA in this context refers to TermAdventure, it can just as well refer to Teaching Assistant.} is implemented as a web application. Based on the comparison in
\cite{crawford2017comparison}, we decided to implement
it in JavaScript using the widely known Node.js runtime dedicated mainly for backend systems. 

There are various types of events that TermAdventure can emit to TA Monitor, namely
\begin{enumerate}
    \item start - the lab has started.
    \item command - user enters an incorrect solution.
    \item passed - user enters a correct solution.
    \item exit - the lab has finished.
    \item help - user raises a hand to call teaching assistant.
    \item ack - teaching assistant approves the help event.
\end{enumerate}

All of the aforementioned events are being emitted using standard UNIX tool called Wget via HTTP POST request. Along
with the event type, the message sent also contains data
such as user name, host name, IP of the computer, ID of
the lab, date, ID of the level and many more. These are used
to identify the request and show the needed information.

\subsection{Solution clustering}
\label{sec:clustering}

The teacher might sometimes want to know what types of
approaches are students using to pass the level. This way
the lecturer is able to see if the solutions used
are matching the expected approach or not.
A standard approach to create groups of similar objects (in our case solutions) is to use a clustering
algorithm. We have chosen the widely known \textit{K-means} \cite{likas2003global}.

One disadvantage of K-means is that user must specify the parameter \textit{K} denoting the number of
groups the solutions are grouped into. This is most of
the times experimental since there is not an exact value
that would work each time. Therefore, the application
allows this number to be specified manually.
The algorithm must be also provided with a distance
function, which is responsible for calculating the distance
between two solutions. Within our application it's possible
to choose between \textit{Jaccard distance} and \textit{Cosine distance} where input vectors are computed as bag of words one-hot encoded vectors.

In order to get a better overview of what these distances look like,
the application is capable of projecting these vectors onto a 2D pane. The
only problem is that our vectors are of high dimension. A machine
learning approach called \textit{t-SNE} \cite{maaten2008visualizing}
is used to perform dimensionality reduction while keeping the objects
that were close in the original space also close in the 2D plane. Such visualization is also useful when dealing with potential plagiarism.

\section{Case Study: UNIX for Data Scientists}

In this section we provide a brief reflection on the use of TermAdventure for a course called ''UNIX for Data Scientists''. Although TermAdventure has been in use for more than five years in various courses at the authors' institution, we chose this particular class for its smaller size (25 students), which also allows for qualitative feedback to be reviewed. Furthermore, most of the class took place remotely due to the COVID-19 pandemic, which provides an opportunity for discussion on how can tools like TermAdventure help with the transition.

The course consists of twelve lectures, which are listed in Table \ref{tab:davos-lectures}. It is an mandatory course for students of the Data Science program and its aim is to help them gain familiarity with the UNIX computing environment. The lectures focus on the basics of UNIX usage and in later lectures the focus shifts onto tools that are of interest to future Data Scientists.

Each lecture is followed by an exercise, organized via TermAdventure. Its goal is to let the students get practical experience with the presented tools and concepts. Almost all of them follow the format discussed in Section \ref{sec:creating-challenges}, with the exception of the sixth class. Its topic is the editor Vim and the exercise for this class involves playing the game of VimGolf: a game in which a specific text change needs to be made with the minimal number of key presses.

The student's experience was very similar to the real world: they used SSH to login to a class server and executed the exercises from there. The server was configured such that each login ended up in a separate \texttt{tmux} session, which allowed the teaching assistants to join the student's session and help them debug their issues. In contrast to the in-person setup, the instructors could not utilize non-verbal clues (like visible signs of frustration) to proactively reach out to struggling students. Hence, monitoring of the class' progress took place almost exclusively via the TA Monitor.

The student's post-term feedback was mostly positive (mean 8.4 out of 10; st. dev 1.3). In free-form responses they praised the fact that it was ''much better than submitting exercises to other e-learning tools'' and mentioned that ''assisted struggle'', in which both the student along with the teaching assistant try to \textit{fight the challenge}, ''is the way to go''. When asked "what could be done better?", the students mostly mentioned bugs in the exercises, which led us to introduce the concept of testing (Section \ref{sec:testing}). As part of their post-term feedback questionnaire, they also voted on the exercise they liked the most and ones that should be improved. The Git one ended up winning the popularity vote, probably due to its immediate practical usefulness. The 10th exercise on Bash scripting received the most negative votes. This is understandable, as in this case we tried to "bend" the framework of TermAdventure to also support the evaluation of Bash scripts. In retrospect, it would have been better to handle this kind of exercise with a specialized tool, such as the one described in \cite{kavspar2019evaluation}.

\begin{table}[]
\begin{tabular}{rlcc}
& Lecture Topic & \textbf{F} & \textbf{I}\\
\hline
1  & Intro to Command Line                   & & $--$ \\
2  & Files and Directories                   \\
3  & Standard I/O, Pipes and Text Processing \\
4  & Processes and Signals                   \\
5  & Users, Groups and Regular Expressions   \\
6  & Vim                                     & ++ \\
7  & File and directory attributes           &&$-$\\
8  & \texttt{find} and \texttt{xargs}                          \\
9  & \texttt{sed} and \texttt{awk}              &&$-$               \\
10 & Intro to Bash scripting         &  & $---$ \\
11 & Git                                    & +++ &  \\
12 & \texttt{csvkit} and \texttt{jq}        &  ++                 
\end{tabular}
    \caption{The twelve lectures which comprise the curricula of the ''UNIX for Data Scientists'' course. The two rightmost columns denote which lectures were the most favorite (\textbf{F}, + signs) and which need the most improvement (\textbf{I}, $-$ signs) according to the results of the end-of-the-term survey.}
    \label{tab:davos-lectures}
\end{table}

\section{Related Work}

With its unceasing popularity as the operating system of machines with large (as well as small) computing power, UNIX and the ability to use it has become part of the standard toolbox of a modern computer scientist. Not only that, due to proliferation of computing tools and significant computing requirements in other areas of science, the knowledge of UNIX has become a requirement even for those who are not trained Computer Scientists, such as data analysts \cite{eckroth2017teaching} or biomedical scientists \cite{mangul2017addressing}.

Ever increasing size of the course audience gave rise to tools that try to effectively teach the command line. All of them aim to facilitate an engaging hands-on experience for the students \cite{simon2005learning}, so that the presented material does not lead to students resenting the command line itself. Some do so by providing automatic evaluation directly inside the command line \cite{solomon2007linuxgym} or by submitting scripts to a web-based autograding solution \cite{kavspar2019evaluation}, while others try to stage the material in form of a treasure hunt \cite{moy2011efficient}.

Some of the current tools require many dependencies, such as servers where solutions are submitted to or they have a need for a webserver, which organizes the class work \cite{kavspar2019evaluation}. This increases the cost, workload on the course staff with setup and maintenance. Moreover, using different environment, such as webpage, to navigate the command line creates the disconnect between student and the command line \cite{bailey2019uassign}. Rather than becoming familiar with the UNIX environment itself, the user needs to make a mental switch when presented to a real command line environment, as they are simply accustomed to a different interface.

In contrast to the previously published works, our solution uses the UNIX command line environment as its sole interaction interface. It utilizes the concept of ''text adventures'' (also referred to as ''interactive fiction''), in which the presented application interactively provides instructions, to which the user replies using a predefined set of commands. It is conceptually close to \texttt{learn}, possibly the first computer-aided instruction program distributed with UNIX itself \cite{kernighan1979learn}, as well as to the concept of self-guided conversational learning \cite{pereira2016leveraging}. Moreover, it is also in line with the suggestion provided by \cite{zachary1994tutorial}, which notes that using conversational style is preferable to the dry style of technical documentation. The ''text adventures'' approach has already been explored in the context of teaching Computer Science for high-school students \cite{proctor2019unfold} but to the best of our knowledge, the presented solution is the first to utilize this concept in the context of teaching UNIX command line.

\section{Conclusions}
In this work we present a suite of tools for creating interactive UNIX exercises under the name TermAdventure (TA). It allows to create exercises in style of text adventure games, which immerse the user in a command line environment and allow them to interact with it. The suite includes TA Monitor that helps a teacher to get an overview of submitted solutions and sends notifications when students get stuck. Moreover, it provides tooling to analyze the student's solutions that can lead to improving the course exercises/structure. The suite  along with documentation and further resources (such as sample challenges) is publicly released under the terms of an open-source license\footnote{\url{https://github.com/NaiveNeuron/TermAdventure}}.



\bibliographystyle{ACM-Reference-Format}
\bibliography{sample-base}

\end{document}